\documentclass[aps,pra,twocolumn,showpacs,floatfix]{revtex4}

\usepackage{amsmath}
\usepackage{txfonts}
\usepackage{microtype}
\usepackage{graphicx}
\usepackage{color}
\usepackage{ulem}

\def\sout{\bgroup\markoverwith
  {\textcolor{red}{\rule[0.5ex]{2pt}{0.5pt}}}\ULon}

\graphicspath{{figures/}}

\usepackage{slashed}
\def\be{\begin{equation}}
\def\ee{\end{equation}}
\def\bes{\begin{equation*}}
\def\ees{\end{equation*}}
\def\bea{\begin{eqnarray}}
\def\eea{\end{eqnarray}}
\def\beas{\begin{eqnarray*}}
\def\eeas{\end{eqnarray*}}
\def\bal#1\eal{\begin{align}#1\end{align}}
\def\bals#1\eals{\begin{align*}#1\end{align*}}

\begin{document}
\title{Above-threshold ionization with highly-charged ions in super-strong\\ laser fields: I. Coulomb-corrected strong field approximation}

\author{Michael Klaiber, Enderalp Yakaboylu, and Karen Z. Hatsagortsyan}
\affiliation{Max-Planck-Institut f\"ur Kernphysik, Saupfercheckweg 1, 69117 Heidelberg, Germany}
\email{klaiber@mpi-hd.mpg.de}

\date{\today}

\begin{abstract}

Aiming at the investigation of above-threshold ionization in super-strong laser fields with highly charged 
ions, we develop a Coulomb-corrected strong field approximation (SFA).
The influence of the Coulomb potential of the atomic core on the ionized electron dynamics in the continuum is taken into account via the eikonal approximation, treating the Coulomb potential perturbatively in the phase of the quasi-classical wave function of the continuum electron. In this paper the formalism of the Coulomb-corrected SFA  for the nonrelativistic regime is discussed employing velocity and length gauge. Direct ionization of a hydrogen-like system in a strong linearly polarized laser field is considered.  The relation of the results in the different gauges to the Perelomov-Popov-Terent'ev imaginary-time method is discussed. 
\end{abstract}

\pacs{32.80.Rm,42.65.-k}

\maketitle

\section{Introduction}
Due to advances in laser technology strong near-infrared laser fields nowadays are available up to intensities of $10^{22}$~W/cm$^2$~\cite{Yanovsky_2008} and much stronger laser fields are envisaged in near future \cite{RMP_2012} stimulating  the investigation of the relativistic regime of laser-atom interaction in ultra-strong fields. The pioneering experiment in this field was carried out by Moore et al. \cite{Moore_1999}. They have investigated the ionization behavior of atoms and ions in a strong laser field  at an intensity of $3\times 10^{18}$ W/cm$^2$. Several further experiments have been devoted to relativistic laser-induced ionization \cite{Chowdhury_2001,Dammasch_2001,Yamakawa_2003,Gubbini_2005,DiChiara_2008,Palaniyappan_2008,DiChiara_2010}.

Numerical investigation of the dynamics of highly-charged ions in super-strong fields has been carried out in \cite{Walser_1999,Hu_1999,Hu_1999b,Casu_2000,Hu_2001,Hu_2002,Walser_2002,Mocken_2004b,Mocken_2008,Hetzheim_2009,Bauke_2011}.
The standard analytical approaches in the field of nonperturbative laser-atom interaction are the strong field approximation (SFA) \cite{Keldysh_1965,Faisal_1973,Reiss_1980} and the imaginary time method (ITM) \cite{Perelomov_1966a,Popov_2005}.
For a theoretical treatment of the relativistic effects, the SFA has been generalized into the relativistic regime in \cite{Reiss_1990,Reiss_1990b} and the ITM in \cite{Popov_1997,Mur_1998,Milosevic_2002r1,Milosevic_2002r2}, respectively. 
In the standard SFA the influence of the Coulomb field of the atomic core is neglected in the electron continuum dynamics and the latter  is described  by the Volkov wave function \cite{Volkov_1935}.
Accordingly, the predictive power of the SFA is the best for negative ions where no long range forces of the parent system act on the ionized electron. For atoms or molecules with long range Coulomb forces the performance of  the SFA downgrades to a qualitative level \cite{Smirnova_2006a}. This is true especially for highly-charged ions.

In the non-relativistic regime the ITM has been successfully used to treat Coulomb field effects during the ionization in the quasi-static regime and the well-known quantitatively correct Perelomov-Popov-Terent'ev (PPT) ionization rate has been derived~\cite{Perelomov_1967a,Ammosov_1986}. The PPT theory uses the quasi-classical wave function for the description of the tunneling part of the electron wave packet through the quasi-static barrier formed by the laser and atomic field, with matching of the  quasi-classical wave function to the exact bound state wave function \cite{Perelomov_1966b,Popov_1967}. The standard SFA technique has also been modified to include Coulomb field effects of the atomic core. The simplest heuristic approach is the, so-called, Coulomb-Volkov ansatz in which the Volkov wave function in the SFA matrix element is replaced by an heuristic Coulomb-Volkov wave function \cite{Jain_Tzoar_1978,Cavaliere_1980,Kaminski_1986,Kaminski_1988,Krstic_1991,Kaminski_1996a,Kaminski_1996b,Ciappina_2007,Yudin_2006,Yudin_2007,Yudin_2008}. In the latter the Coulomb field is taken into account via an incorporation of the asymptotic phase of the exact Coulomb-continuum wave function into the phase of the heuristic Coulomb-Volkov wave function \cite{Faisal_2008}. Consequently, the coupling between the Coulomb and laser field is neglected in the Coulomb-Volkov ansatz and the approach fails when the electron appears in the continuum after tunneling close to the atomic core~\cite{Smirnova_2006b}.

Following a more rigorous approach, the eikonal approximation \cite{Glauber_1959} has been proposed to apply for strong field problems \cite{Gersten_1975}. In the latter, nonrelativistic free-free transitions in the laser and the Coulomb field have been considered employing an eikonal wave function for the continuum electron. Here the laser field is taken into account exactly, while the Coulomb field is via the eikonal approximation. The eikonal approximation has been generalized in
 \cite{Avetissian_1997} to include quantum recoil effects at photon emission and absorption. A Coulomb-corrected SFA for nonrelativistic ionization employing the eikonal wave-function has been first proposed in \cite{Krainov_1997}. Similar approaches have been considered in \cite{Krainov_1995,Gordienko_2003,Goreslavski_2004,Faisal_2005,Faisal_2006,Chirila_2005}. Recently, the nonrelativistic Coulomb-corrected SFA based on the eikonal-Volkov wave function for the  continuum electron  has been further elaborated  in \cite{Smirnova_2007,Smirnova_2008} and applied for molecular strong field ionization and high-order harmonic generation. The Coulomb-corrected SFA has also been extended to include rescattering effects \cite{Popruzhenko_2008a,Popruzhenko_2008b}. Here the Coulomb field is taken into account exactly in the quasi-classical electron continuum trajectories that are later plugged  into the phase of the quasi-classical wave function.

In the relativistic regime, similar to the nonrelativistic case, the standard SFA is only exponentially exact since the Coulomb field is neglected during  ionization, whereas the ITM \cite{Popov_1997,Mur_1998,Milosevic_2002r1,Milosevic_2002r2} can provide also correct preexponential factors. Can the quantitatively correct relativistic ionization probabilities be derived via  the SFA technique accounting Coulomb field effects accurately?  The relativistic generalized eikonal-Volkov wave function (taking also into account quantum recoil) has been derived in \cite{Avetissian_1999}. The Coulomb corrected SFA based on this wave function has been proposed in \cite{Avetissian_2001}. However, final results have been obtained only in Born approximation, i.e. via an expansion of the eikonal wave function with respect to the Coulomb field, which, in fact, reduces the transition matrix element to the one  in the standard second order SFA.

With this paper we  begin a sequel of papers in which we develop the relativistic Coulomb-corrected SFA based on the Dirac equation, generalizing the nonrelativistic theory of \cite{Krainov_1997,Smirnova_2008} and apply it for the calculation of spin-resolved quantitatively correct ionization probabilities. Rather than the Volkov wave function, the eikonal-Volkov wave function is employed as final state of  the Coulomb corrected SFA. The influence of  Coulomb potential of the atomic core on the ionized electron continuum dynamics is taken into account via the eikonal approximation. The latter means that the quasi-classical (WKB) approximation is applied for the electron continuum dynamics and, additionally, the Coulomb potential is treated perturbatively in the phase of the quasi-classical wave function. The formalism is applied for direct ionization of a hydrogen-like system in a strong linearly polarized laser field.

In this first paper of the sequel, we begin with the nonrelativistic Coulomb-corrected SFA to show in the  most simple case the scheme of the Coulomb-corrected SFA. The SFA formalism is applied to treat the Coulomb field effect of the atomic core during  ionization systematically and to obtain quantitatively correct results which, in particular, for the total ionization rate coincide with the PPT result. Two versions of the theory based on the velocity and length gauge, respectively, are considered. Comparison with the PPT theory is carried out and the physical relevance of the two versions is discussed. A conclusion is drawn concerning the scheme of the relativistic generalization of the Coulomb-corrected SFA. In the second paper of the sequel, the relativistic Coulomb-corrected SFA will be developed, and the next paper in the sequel will be devoted to spin effects in relativistic above-threshold ionization.

The plan of the paper is the following: In section~\ref{NCCSFA} the nonrelativistic Coulomb-corrected SFA in the length gauge is considered and differential and total ionization rates for hydrogen-like systems are derived. The next section is dedicated to the   Coulomb-corrected SFA in velocity gauge.  The comparison of the different versions of the Coulomb-corrected SFA is carried out in Sec.~\ref{Comparison}, and the conclusion is given in Sec.~\ref{Conclusion}.

\section{Nonrelativistic Coulomb-corrected SFA in the length gauge}\label{NCCSFA}

In this section we show how the nonrelativistic Coulomb-corrected SFA in the length gauge is developed. Rather than the usual Volkov wave function, it employs the eikonal-Volkov wave function to describe the electron continuum dynamics accurately, taking into account the Coulomb field effect of the atomic core. As we will see  in this way the PPT ionization rates can be recovered within the SFA formalism.

\subsection{The standard SFA}\label{cSFA}

We consider a highly-charged hydrogen-like ion interacting with a laser field. The dynamics is governed by the Hamiltonian 
\begin{eqnarray}
  H=H_0+H_{int},
\label{Hamiltonian}
\end{eqnarray}
where $H_0$ is the Hamiltonian of the atomic system
\begin{eqnarray}
  H_0=\hat{\mathbf{p}}^2/2+V(\mathbf{r}),
\end{eqnarray}
with the atomic potential $V(\mathbf{r})$, the momentum operator $\hat{\mathbf{p}}$ and coordinate vector $\mathbf{r}$ (atomic units are used throughout). The interaction Hamiltonian due to the laser field in length gauge is
\begin{eqnarray}
 H_{int}(t)=\mathbf{r}\cdot\mathbf{E}(t),
\end{eqnarray}
with the laser electric field $\mathbf{E}(t)$.
The time evolution operator $U(t,t_0)$ of the atom in the laser field can be formulated via the Dyson equation
\begin{eqnarray}
  U(t,t_0)=U_0(t,t_0)-i\int^t_{t_0} dt U(t,t') H_{int}(t')U_0(t',t_0)
\end{eqnarray}   
where $U_0$ is the time evolution operator of the atomic system without the laser field. The matrix element for a laser induced transition from the initial atomic ground state $|\phi(t)\rangle=|0\rangle e^{i I_pt}$, with the ground state energy $-I_p$, and the ionization potential $I_p\equiv \kappa^2/2$,
into a continuum eigenstate of the total system $|\psi_{\mathbf{p}}(t)\rangle$ with an asymptotic momentum $\mathbf{p}$ is then given by
\begin{eqnarray}
  M_{\mathbf{p}}=-i\int^{\infty}_{-\infty} dt \langle\psi_{\mathbf{p}}(t)|H_{int}(t)|\phi(t)\rangle.
\label{M_SFA}
\end{eqnarray}
In the SFA, the final continuum state is approximated by a Volkov state $|\psi_{\mathbf{p}}^V(t)\rangle$, i.e. an eigenstate of a Hamiltonian, where the electron is only interacting with the laser field \cite{Volkov_1935}. In coordinate space it is given by
\begin{eqnarray}
  \langle \mathbf{r}  |\psi_{\mathbf{p}}^V(t)\rangle =\exp[i S^{(0)}_0(\mathbf{r},t)]/(2\pi)^{3/2}.
\end{eqnarray}
The function in the exponent $S^{(0)}_0(\mathbf{r},t)=(\mathbf{p}+\mathbf{A}(t))\cdot \mathbf{r}+\int^{\infty}_{t}dt'\left(\mathbf{p}+\mathbf{A}(t')\right)^2/2$ is the classical action of an electron in a laser field in the length gauge. Note that the Volkov wave function coincides exactly with the wave function in the zeroth-order WKB approximation for the system. The ionization matrix element in the SFA yields:
\begin{eqnarray}
  M_{\mathbf{p}}=-i\int^{\infty}_{-\infty} dt \langle \mathbf{p}+\mathbf{A}(t)|H_{int}(t)|0\rangle \exp[-i\tilde{S}(t)]
  \label{Mfi}
\end{eqnarray}
with $\tilde{S}(t)=\int^{\infty}_t dt'[(\mathbf{p}+\mathbf{A}(t'))^2/2+\kappa^2/2]$.
In the adiabatic regime, when the laser frequency $\omega$ is smaller than the ground state energy $I_p$ and the ponderomotive potential $U_p=E_0^2/4\omega^2$, with the laser field amplitude $E_0$, the time integration in Eq. (\ref{Mfi}) can be carried out in good accuracy via the saddle point method (SPM), see, e.g.,~\cite{Gribakin_1997}. This yields   
\begin{eqnarray}
  M_{\mathbf{p}}=-i\sum_{s}\sqrt{\frac{2\pi}{i\ddot{\tilde{S}}(t_s)}}\langle \mathbf{p}+\mathbf{A}(t_s)|H_{int}(t_s)|0\rangle \exp[-i\tilde{S}(t_s)],
  \label{MS}
\end{eqnarray}
where $t_s$ are the so-called saddle points of the integrable function 
defined by $\dot{\tilde{S}}(t_s)=0$. 
After a partial integration in Eq. (\ref{Mfi}),  the transition operator in the matrix element can be transformed from $H_{int}$ to $V(\mathbf{r})$~\cite{Becker_2002}: 
\begin{eqnarray}
  M_{\mathbf{p}}=-i\sum_{s}\sqrt{\frac{2\pi}{i\ddot{\tilde{S}}(t_s)}}\langle \mathbf{p}+\mathbf{A}(t_s)|V(\mathbf{r})|0\rangle \exp[-i\tilde{S}(t_s)].
  \label{MS2}
\end{eqnarray}
In the case of a long laser pulse the differential ionization rate is expressed via the matrix element as follows 
\cite{Gribakin_1997}:
\begin{eqnarray}
  \frac{dw}{d^3\mathbf{p}}&=&\frac{\omega}{2\pi}|M_{\mathbf{p}}|^2,
  \label{io}
\end{eqnarray}
where the summation in Eq.~(\ref{MS}) is carried out only over the saddle points of one laser period.

\subsection{SFA for a negative ion}

The calculation of ionization rates is straightforward in the case of ionization of a negative ion. 
The latter can be modeled by a zero-range potential $V^{(z)}(\mathbf{r})=-(2\pi/\kappa)\delta(\mathbf{r})\partial_r r$, with the matrix element $\langle\mathbf{p}|V^{(z)}|0^{(z)}\rangle=-\sqrt{\kappa}/(2\pi)$~\cite{Milosevic_2002c}. 
In a sinusoidal laser field $\mathbf{A}(t)=(\mathbf{E}_0/\omega)\sin(\omega t)$ the saddle point equation yields:
\begin{eqnarray}
  \sin(\omega t_s)=-\frac{p_E}{E_0/\omega}+i \sqrt{\gamma^2+\left(\frac{p_{\perp}}{E_0/\omega}\right)^2}
  \label{spc}
\end{eqnarray} 
with the Keldysh parameter $\gamma\equiv \kappa\omega/E_0$, $p_E\equiv \mathbf{p}\cdot\hat{\mathbf{e}}$, $p_{\perp}\equiv |\mathbf{p}-(\mathbf{p}\cdot\hat{\mathbf{e}})\,\hat{\mathbf{e}}|$, $\hat{\mathbf{e}}\equiv  \mathbf{E}_0/|\mathbf{E}_0|$. In the tunneling regime ($\gamma\ll 1$)
the saddle points in one laser cycle can be given approximately via a perturbative solution of Eq.~(\ref{spc}) with respect to $\gamma$:
\begin{eqnarray}\
  \omega t_{s1}&=&-\arcsin\left[\frac{p_E}{E_0/\omega}\right]+i \frac{\sqrt{\kappa^2+p_{\perp}^2}}{|E(t_0)|/\omega}\nonumber\\
  \omega t_{s2}&=&\pi +\arcsin\left[\frac{p_E}{E_0/\omega}\right]+i \frac{\sqrt{\kappa^2+p_{\perp}^2}}{|E(t_0)|/\omega}.
\end{eqnarray}
with $|E(t_0)|= E_0\sqrt{1-\left(\omega p_E/E_0\right)^2}$. 
Inserting  the saddle points into Eq.(\ref{io}), yields the differential ionization probability of a negative ion
\begin{eqnarray}
  \frac{dw^{(z)}}{d^3\mathbf{p}}= \frac{\omega}{2\pi^2|E(t_0)|}\exp\left[-\frac{2\left(\kappa^2+p_{\perp}^2\right)^{3/2}}{3|E(t_0)|}\right]
  \label{io2}
\end{eqnarray}
Since the ratios $p_E/(E_0/\omega)$ and $p_\perp/(E_0/\omega)$ are smaller than one in the case of tunnel ionization, we can expand the function in the exponent quadratically in terms of momentum and neglect the dependence in the preexponential factor. With this we arrive at the differential ionization rate:
\begin{eqnarray}
\label{z_diff_rate}
  \frac{dw^{(z)}}{d^3\mathbf{p}}= \frac{\omega}{2\pi^2E_0}\exp\left[-\frac{2\kappa^3}{3E_0}-\frac{\kappa}{E_0}p_{\perp}^2-\frac{\kappa^3\omega^2}{3E_0^3} p_E^2\right],
\end{eqnarray}
and the total ionization rate:
\begin{eqnarray}
\label{z_total_rate}
  w^{(z)}=\sqrt{\frac{3}{\pi}}\frac{E_0^{3/2}}{2\kappa^{5/2}}\exp\left[-\frac{2\kappa^3}{3E_0}\right].
\end{eqnarray}
The SFA ionization rates of Eqs. (\ref{z_diff_rate}) and (\ref{z_total_rate}) for a short range potential coincide with the ITM result~\cite{Popov_2004}. The physical reason is that neglecting the atomic potential after the electron is transferred into the continuum, is justified for negative ions.

\subsection{SFA for a hydrogen-like system}\label{SFACP}

In the case of atomic ionization, the Coulomb potential of the ionic core cannot be neglected in the electron continuum dynamics. Therefore, to obtain an accurate ionization rate, the wave function of the continuum state $|\psi_{\mathbf{p}}(t)\rangle$ in the SFA ionization amplitude is approximated  by the eikonal wave function (instead of the usual Volkov function) which accounts for the Coulomb field effect of the ionic core \cite{Krainov_1997,Smirnova_2008}.

As we noted in the previous section, the Volkov wave function is identical to the electron wave function in the laser field in the zeroth order WKB-approximation.
A systematic improvement of this state compared to the exact continuum state can be achieved employing 
the WKB-approximation for the wave function of an electron exposed to the simultaneous action of the laser  and the Coulomb field. From the Schr\"odinger equation for an electron in a Coulomb potential $V^{(c)}(\mathbf{r})=-\kappa/r$ and a laser field $\mathbf{E}(t)$
\begin{eqnarray}
  i\hbar \partial_t\psi=-\dfrac{\hbar^2}{2}\Delta \psi +V^{(c)}\psi+\mathbf{r}\cdot\mathbf{E}(t)\psi,
\end{eqnarray}
the ansatz $\psi=e^{i S/\hbar}$ yields the following equation
\begin{eqnarray}
  -\dot{S} = \frac{(\boldsymbol{\nabla}S)^2}{2}+ V^{(c)} +\mathbf{r}\cdot\mathbf{E}+ \dfrac{\hbar}{i}\frac{\Delta S}{2}.
\end{eqnarray}
Using the WKB-expansion $S=S_0 + \frac{\hbar}{i}S_1 + \ldots$, we obtain the equation
\bea
&& \left(\dfrac{\hbar}{i}\right)^0: \quad -\dot{S}_0 = \dfrac{\left(\boldsymbol{\nabla}S_0\right)^2}{2} + V^{(c)}+\mathbf{r}\cdot\mathbf{E}(t), 
\eea
$S_0$ is the classical action of an electron in the laser field and the atomic potential. In the eikonal approximation the partial differential equation for $S_0$ is solved perturbatively in
the atomic potential $V^{(c)}$. The zeroth order solution gives the Volkov-action 
\begin{eqnarray}
S^{(0)}_0 (\mathbf{r},t)=(\textbf{p}+\textbf{A}(t))\cdot\textbf{r}+\frac{1}{2}\int_t^{\infty} dt'\left( \textbf{p}+\textbf{A}(t')\right)^2,
\end{eqnarray}
with $\textbf{A}(t)=-\int_{-\infty}^t dt' \textbf{E}(t')$, whereas the first order solution reads
\begin{eqnarray}
  S^{(1)}_0(\mathbf{r},t)
  =\int^{\infty}_t dt' V^{(c)}\left(\mathbf{r}(t')\right),
  \label{S1}  
\end{eqnarray}
with the trajectory of the electron in the laser field $\mathbf{r}(t')=\mathbf{r}+\int^{t'}_{t}dt''\mathbf{p}(t'')$ and $\mathbf{p}(t)\equiv \mathbf{p}+\mathbf{A}(t)$.
The time $t$ can be interpreted as the time and $\mathbf{r}$ as the coordinate of the ionization event. 
Thus, the approximate wave function of the electron continuum state in the laser and Coulomb field, which is termed as the eikonal-Volkov wave function, in the nonrelativistic regime is
\begin{eqnarray}
\psi_{\mathbf{p}}^{(c)}(\mathbf{r},t)=\frac{1}{(2\pi)^{3/2}}\exp\{i S_0^{(0)}(\mathbf{r},t)+i S_0^{(1)}(\mathbf{r},t)\}.
\end{eqnarray} 
It takes into account the influence of the atomic potential quasi-classically up to first order and will be used in the SFA amplitude of Eq. (\ref{M_SFA}).

Let us estimate the applicability of the eikonal approximation given by the condition $S^{(1)}_0\ll \tilde{S}$. The perturbed action can be estimated $S^{(1)}_0 \sim \int V^{(c)}d\tau\sim \int d\tau\dot{x}/x\sim\log(r_{E\,e}/r_{E\,i})\sim\log(\sqrt{E_a/E_0})\sim 1$, using the potential $V^{(c)}\sim \kappa/r_E$ ($r_E\equiv \textbf{r}\cdot \hat{\textbf{e}}$), the initial coordinate before tunneling $r_{E\,i}\sim v_c \delta\tau_c$, the velocity $v_c\sim \kappa$, the uncertainty of the initial time $\delta\tau_c$ [in the latter, we use the time-width of the saddle-point integration $\delta\tau_c\sim 1/\sqrt{\ddot{\tilde{S}}(t_s)}\sim1/\sqrt{\kappa E_0}$] and the tunnel exit coordinate $r_{E\,e}\sim\kappa^2/E_0$. While the Volkov-action is estimated  $\tilde{S} \sim p(\tau_c)^2\tau_c+I_p\tau_c\sim E_0^2\tau_c^3+I_p\tau_c\sim E_a/E_0$ with the tunneling time $\tau_c\sim \gamma/\omega=\kappa/E_0$ determined by the Keldysh parameter and the atomic field $E_a=\kappa^3$, the eikonal approximation for the nonrelativistic ionization problem is valid when
\begin{eqnarray}
  \frac{E_0}{E_a}\ll 1.
  \label{eikonal-condition}
\end{eqnarray}
Note that $E_0/E_a<1/16$ in the tunneling ionization regime for a hydrogen-like ion.

To be able to handle the additional term $S_0^{(1)}$ in the SFA transition amplitude, we have to make simplifications. The time derivative of $S_0^{(1)}$ given by $\partial_t{S}_0^{(1)}(\mathbf{r},t)\approx-V^{(c)}\left(\mathbf{r}+\int^{\infty}_{t}dt''\mathbf{p}(t'')\right)$, corresponds to the potential energy of the ionized electron in the remote future, after it has escaped from the bound state.
Since the electron left the atomic system after ionization and recollision is not considered here, its potential energy is vanishing for asymptotically large times and therefore it is justified to use $\dot{S}_0^{(1)}(\mathbf{r},t)\approx 0$. Consequently, the additional term $S_0^{(1)}$ in the exponent of the amplitude has no influence on the saddle point equation and leaves the saddle points unchanged \footnote{This is in contrast to  \cite{Popruzhenko_2008a,Popruzhenko_2008b}, where recollisions of the ionized electron are considered and the Coulomb field influence on the electron trajectory after the liberation from the atom is explicitly taken into account.}, however, it can change the preexponential by a factor $\exp[-i S_0^{(1)}(\mathbf{r},t_s)]$.
Thus, the Coulomb-corrected SFA amplitude of ionization reads: 
\begin{eqnarray}
  M^{(c)}_{\mathbf{p}}&=&-i\int^{\infty}_{-\infty} dt \langle \mathbf{p}+\mathbf{A}(t)|H_{int}(t)\exp[-i S_0^{(1)}(\mathbf{r},t)]|0^{(c)}\rangle \nonumber\\
  &&\times\exp\left[-i\tilde{S}(t)\right],
  \label{Mc}
\end{eqnarray}
where $|0^{(c)}\rangle$ is the electron bound state in the Coulomb potential. The next task is to find an analytic expression for the new preexponential factor for times $t=t_s$. 
Physically, $S_0^{(1)}(\mathbf{r},t)$ corresponds to the sum of potential energies the electron possesses on its trajectory. When the electron has left the vicinity of the  atomic core, the potential energy is small and there are no further contributions to $S_0^{(1)}(\mathbf{r},t)$. Since we consider the tunneling regime where $E_0/\omega\gg \kappa$, this situation sets in at a very moment of ionization. Thus, it is justified to expand the argument in $S_0^{(1)}$ describing the trajectory of the electron, up to second order around the saddle point $t_s$, i.e. around the instant of ionization:
\begin{eqnarray}
  \mathbf{r}(t')=\mathbf{r}+\mathbf{p}(t_s)(t'-t_s)-\mathbf{E}(t_s)(t'-t_s)^2/2.
\end{eqnarray}
Further, the momentum distribution of the amplitude is dominated by the exponential function that is located around the laser polarization direction, i.e. we can assume in the preexponential function $\mathbf{p}=p_E \hat{\mathbf{e}}$ and $\mathbf{p}(t_s)=i\kappa\hat{\mathbf{e}}$. Additionally, it can be argued that the tunnel ionization starts mainly  in the area around the laser polarization axis $\mathbf{r}=r_E\hat{\mathbf{e}}$, i.e. at the outskirts of the atom in direction of the laser electric field. This typical value for the initial coordinate of the trajectory $\mathbf{r}$ is justified via the saddle point condition for the integral $\int d^3\mathbf{r}\exp[-i\mathbf{p}(t_s)\cdot\mathbf{r}-\kappa r]$ which leads to $\mathbf{r}_s/r_s=\mathbf{p}(t_s)/(i\kappa)$. Thus, the integrand in the expression of the Coulomb-correction factor of Eq. (\ref{S1}) can  be simplified:
\begin{eqnarray}
  \frac{1}{r(t')}&=&\frac{1}{\left|r_E+p_E(t_s)(t'-t_s)-E(t_s)(t'-t_s)^2/2\right|}. 
\end{eqnarray}
Furthermore, the motion after the electron has left the barrier, contributes only as an unimportant phase in the preexponential factor in Eq.~(\ref{Mc}) and the integration limit can be set at the tunnel exit: $\omega t_0 =-\arcsin\left[p_E/(E_0/\omega)\right]$. With these simplifications the integral in Eq. (\ref{S1}) can be evaluated:
\begin{eqnarray}
  \exp\left[-i S_0^{(1)}(\mathbf{r},t)\right]&=&\left(\frac{1 + \sqrt{1 + 4 \lambda}}{-1 + \sqrt{1 + 4 \lambda}}\right)^{\frac{1}{\sqrt{1 + 4 \lambda}}}
  \nonumber\\
  &\approx&\frac{1}{\lambda}+{\rm O}(\lambda),
\end{eqnarray}
with the small quantity $\lambda=-\mathbf{r}\cdot\mathbf{E}(t_s)/2\kappa^2$ which is of the order of $\sqrt{E_0/E_a}\ll 1$, see Eq.~(\ref{eikonal-condition}). In fact, one can estimate $\lambda\sim x_c E_0/\kappa^2\sim v_a\tau_c E_0/\kappa^2\sim \sqrt{E_0/E_a}$. 
We underline that in all expressions after Eq.~(\ref{eikonal-condition}) expansions in this parameter are employed.

We come to the conclusion that in the nonrelativistic regime the Coulomb-corrected SFA amplitude differs from the one in the standard SFA by the following Coulomb-correction factor:
\begin{eqnarray}
  Q_{nr}=-\frac{4I_p}{\mathbf{r}\cdot\mathbf{E}(t_s)}.
\end{eqnarray}
The transition amplitude can then be expressed in a very simple form:
\begin{eqnarray}
  M^{(c)}_{\mathbf{p}}=4i I_p \int^{\infty}_{-\infty} dt \langle \mathbf{p}+\mathbf{A}(t)|0^{(c)}\rangle \exp\left\{-i\tilde{S}(t)\right\}.
  \label{mc}
\end{eqnarray}
This simple form for the ionization amplitude in length-gauge Coulomb-corrected SFA is achieved because the Coulomb-correction factor $Q_{nr}$ cancels the dipole interaction factor $\textbf{r}\cdot \textbf{E}$ in the length-gauge  matrix element.

The occurring matrix element is singular at the saddle point: 
\begin{eqnarray}
  \langle\mathbf{p}+\mathbf{A}(t)|0^{(c)}\rangle& =&\frac{1}{\pi}\frac{2 \sqrt{2} \kappa ^{5/2}}{  \left[\kappa ^2+(\mathbf{p}+\mathbf{A}(t))^2\right]^2}\nonumber\\
  &=&-\sqrt{\frac{\kappa}{2}} \frac{1}{\pi\mathbf{E}(t_s)^2(t-t_s)^2},
\end{eqnarray}
where in the last step only the leading order term in $E_0/E_a$ is retained, and the integral in Eq.~(\ref{mc}) must be calculated via the modified SPM~\cite{Gribakin_1997}, taking into account the pole during the integration. Compared to the case of a zero-range potential  this yields a correction factor in the amplitude of
\begin{eqnarray}
  \frac{M^{(c)}}{M^{(z)}}=\frac{2^{3/2}E_a}{|E(t_0)|}.
\end{eqnarray}
This correction factor is known from ITM \cite{Perelomov_1967a} but appears to be reproducible also with the SFA technique. The differential ionization rate in the case of a Coulomb potential of the atomic core is 
\begin{eqnarray}
  \frac{dw^{(c)}}{d^3\mathbf{p}}= \frac{4}{\pi^2}\frac{\omega\kappa^6}{E_0^3}\exp\left[-\frac{2E_a}{3E_0}-\frac{\kappa}{E_0}p_{\perp}^2-\frac{\kappa^3\omega^2}{3E_0^3} p_E^2\right],
  \label{dwdp_nr}
\end{eqnarray}
and the total ionization rate yields
\begin{eqnarray}
  w^{(c)}&=&4\sqrt{\frac{3}{\pi}}\frac{\kappa^{7/2}}{E^{1/2}_0}\exp\left[-\frac{2\kappa^3}{3E_0}\right].
  \label{w_c}
\end{eqnarray}
These rates are identical to the PPT-ionization rate \cite{Perelomov_1967a,Ammosov_1986}.
The momentum distribution of the ionized electrons in the non-relativistic regime indicates that the emission of electrons with a vanishing final momentum is most probable. The longitudinal and the transversal widths of the distribution are $\Delta_{\parallel}=\sqrt{E_0/E_a}E_0/\omega$ and  $\Delta_{\bot}= \sqrt{E_0/E_a}\kappa$, respectively.

Concluding this section, within the SFA S-matrix formalism and employing the eikonal-Volkov wave function for the description of the laser-driven electron continuum dynamics disturbed by the atomic Coulomb potential, as well as neglecting recollisions, one can derive quantitatively correct differential as well as total ionization rates that coincide with the expressions obtained within the PPT quasi-static theory. In the next section we apply the Coulomb-corrected SFA formalism in velocity gauge.

\section{Nonrelativistic Coulomb-corrected SFA in velocity gauge}\label{NCCSFA2}

It is well known that the SFA is, in general, not gauge-invariant and the SFA in different gauges correspond to  different physical approximations. In this section we calculate the ionization rate of an hydrogen-like ion using the Coulomb-corrected SFA in velocity gauge. Later, we will compare it with the results of the PPT theory and the length-gauge Coulomb-corrected SFA to answer the question: in which gauge the Coulomb-corrected SFA is more relevant for the calculation of the ionization rate of an hydrogen-like ion? We will use this information in the next paper for the development of the relativistic Coulomb-corrected SFA.

In velocity gauge the Hamiltonian is given by Eq. (\ref{Hamiltonian}) with the interaction Hamiltonian 
\begin{eqnarray}
  H_{int}(t)=\mathbf{p}\cdot\mathbf{A}(t)+\mathbf{A}(t)^2/2.
\end{eqnarray}
The corresponding Volkov wave function describing the free electron in the laser field in this gauge is
\begin{eqnarray}
  \psi^V(\mathbf{r},t)=\frac{1}{\sqrt{2\pi}^3}\exp[i\mathbf{p}\cdot\mathbf{r}+i\tilde{S}(t)].
\end{eqnarray} 
In the case of ionization of a negative ion, the ionization amplitude in the standard SFA in velocity gauge is given by Eq. (\ref{MS2}) where the preexponential matrix element is replaced:
\begin{eqnarray}
   \langle\mathbf{p}+\mathbf{A}(t)|V|0\rangle\,\,\ \rightarrow \langle\mathbf{p}|V|0\rangle.
\end{eqnarray}
Since the matrix element $\langle\mathbf{p}|V|0\rangle$ is constant and does not depend on momentum in the case of a short-range potential, it is identical to the one in the length gauge.  
Therefore, the overall ionization amplitude for a negative ion is gauge-invariant in the standard SFA.

In the case of a Coulomb-potential as ionic core the situation is different. Here the preexponential matrix-element is not a constant and the different momentum dependencies could lead to a gauge dependence. 
The Coulomb corrected SFA based on the eikonal-Volkov solution can be developed for the velocity gauge similar to that in the previous section. The same steps lead to
the following final expression for the ionization amplitude, cf. Eq.~(\ref{mc}),
\begin{eqnarray}
  M^{(c)}_{\mathbf{p}}&=&-i\int^{\infty}_{-\infty} dt \langle \mathbf{p}|Q_{nr}\left[\mathbf{p}\cdot\mathbf{A}(t)+\mathbf{A}(t)^2/2\right]|0^{(c)}\rangle
  \nonumber\\
  &&\times\exp\left\{-i\tilde{S}(t)\right\}.
  \label{mc2}
\end{eqnarray}
In contrast to the length gauge calculation, the saddle point of $\tilde{S}$ lays not on the singularity of the preexponential matrix element and the standard saddle point approximation can be applied. It yields for the amplitude
\begin{eqnarray}
  M^{(c)}_{\mathbf{p}}&=&\frac{2 E_a \left[p_E (p_E - 2 i \kappa) - 
   2 (p_E - i \kappa) \kappa \arctan\left(\frac{p_E}{\kappa}\right)\right]}{\sqrt{\pi}p_E^2 |E(t_0)|^{3/2}}\nonumber\\
 &&\times\exp\left[-\frac{\left(\kappa^2+p_{\perp}^2\right)^{3/2}}{3|E(t_0)|}\right].
\end{eqnarray}
The  ionization differential rate in the velocity gauge Coulomb-correct SFA reads
\begin{eqnarray}
&&\frac{dw^{(c)}}{d^3\mathbf{p}}=\frac{4 \kappa ^6 \omega }{\pi ^2 E_0^3}\exp\left[-\frac{2\kappa^3}{3E_0}-\frac{\kappa}{E_0}p_{\perp}^2-\frac{\kappa^3\omega^2}{3E_0^3} p_E^2\right]\nonumber\\
&&\times\left\{1+\frac{4\kappa^2}{p_E^2}-\frac{4 \kappa }{p_E}\arctan\left(\frac{p_E}{\kappa }\right)\left(1 +\frac{2\kappa^2}{p_E^2}\right)\right.\nonumber\\
&&+\left.\frac{4\kappa^2}{p_E^2}\arctan\left(\frac{p_E}{\kappa }\right)^2\left(1+ \frac{\kappa^2}{p_E^2}\right)\right\}.
\label{VG_diff_rate}
\end{eqnarray}
The ionization  differential rate in the velocity gauge differs from that in the length gauge, see Eq.~(\ref{dwdp_nr}), by the expression in the curly brackets in Eq.~(\ref{VG_diff_rate}).
To obtain the total ionization rate, the $p_{\bot}$-integration can be carried out analytically, but  $p_E$-integration has to be accomplished numerically.

\begin{figure}
  \begin{center}
    \includegraphics[width=0.4\textwidth]{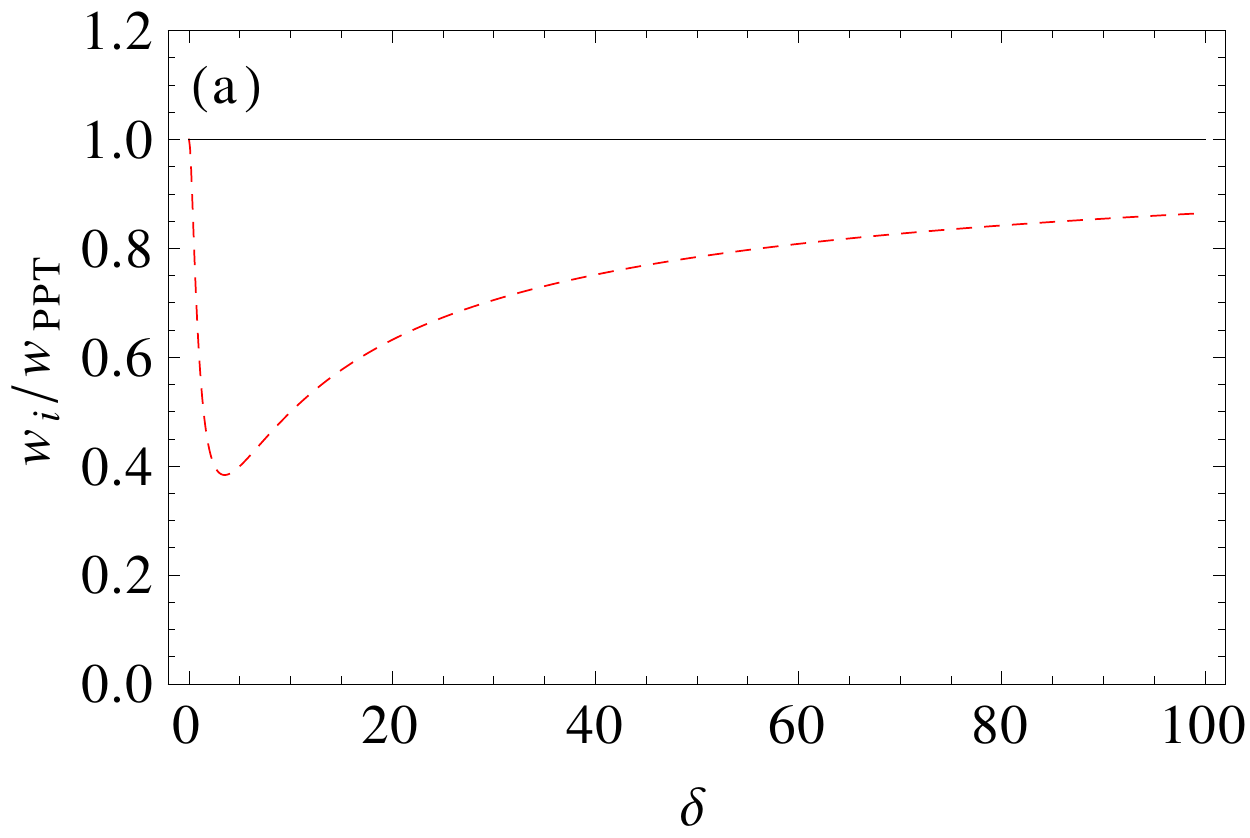}
\includegraphics[width=0.4\textwidth]{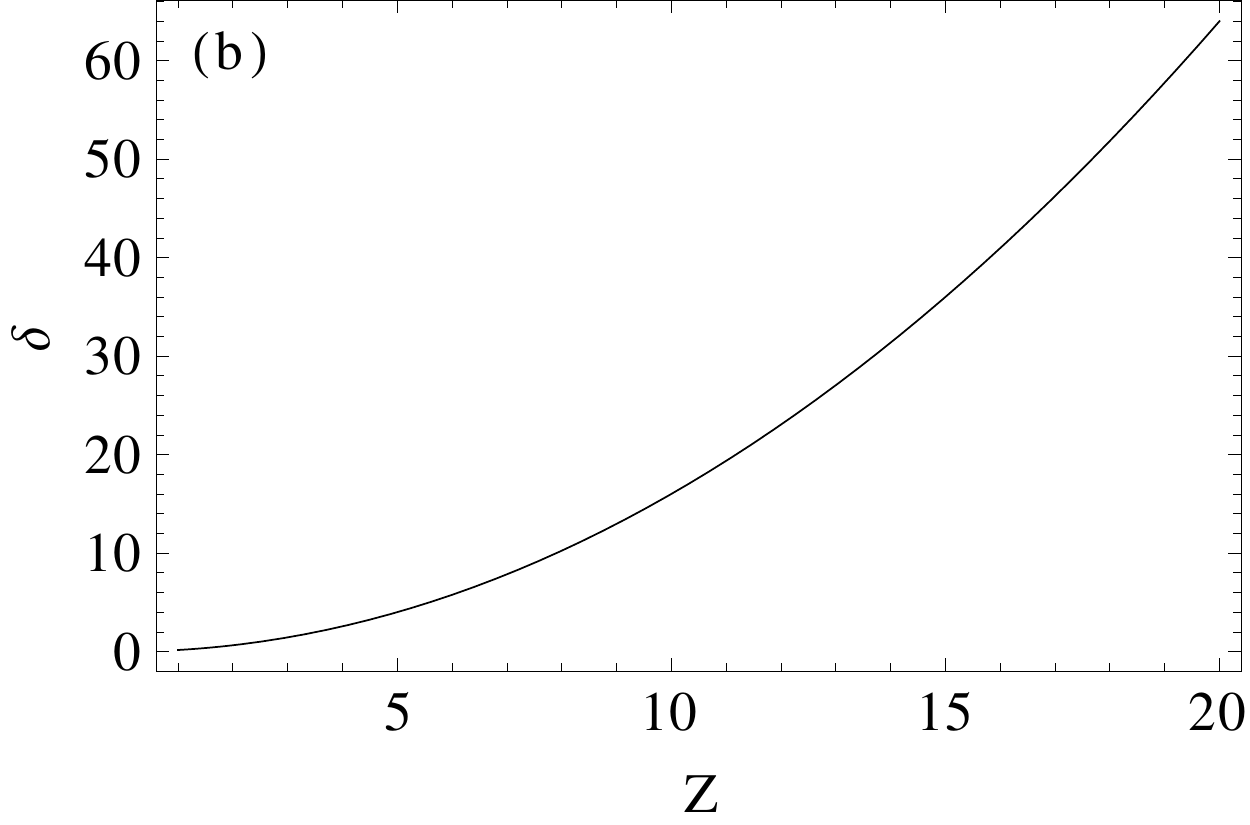}
       \caption{(a) The ratio of the total ionization rate derived in the Coulomb-corrected SFA with respect to the PPT rate vs the parameter $\delta=\sqrt{E_0 /E_a}/\gamma$: SFA in length gauge (black, solid), and SFA in velocity gauge (red, dashed); (b) The parameter $\delta$ for different nuclear charges $Z$ at a fixed angular frequency $\omega=0.05$ a.u. and $E_0/E_a=1/25$. The laser intensity is $I=5.6\times 10^{19}\times (Z/10)^6$ W/cm$^2$.}
    \label{ADKVG}
  \end{center}
\end{figure}

In Fig.~\ref{ADKVG}~(a) we compare the total ionization rate calculated within the Coulomb-corrected SFA in the length or velocity gauge with the PPT rate for different values of the parameter $\delta=\sqrt{E_0/E_a}/\gamma=(E_0/E_a)^{3/2}(I_p/\omega)$. This parameter arises since the deviation in the two gauges depends on the curly bracket that is a function of $p_E/\kappa$ with the typical value for the momentum in laser polarization direction $p_E\sim \Delta_{\parallel}=\sqrt{E_0 /E_a}E_0/\omega$. 
While the length-gauge result coincides with the PPT one, the velocity gauge results tends to the PPT-rate  only in the limits $\delta\rightarrow 0$ \cite{Krainov_1997} and   $\delta\rightarrow \infty$, deviating from the latter at intermediate values of $\delta$. This is evident from Eq. (\ref{VG_diff_rate}), since the curly bracket goes to one in both limits. For intermediate values of the parameter $\delta$, the deviation can be larger than a factor of 2. Note that in the tunneling regime the parameter $\delta$ can vary in the total range of $(0,\infty)$. In Fig.~\ref{ADKVG}~(b) we show the value of $\delta$ for different nuclear charges $Z$ and a suboptical angular frequency. It can be seen that for this parameter set the value of $\delta$ lays in an area where the results in the two gauges differ significantly.   

\section{Comparison of different approximations}\label{Comparison}

\begin{figure}
  \begin{center}
    \includegraphics[width=0.4\textwidth]{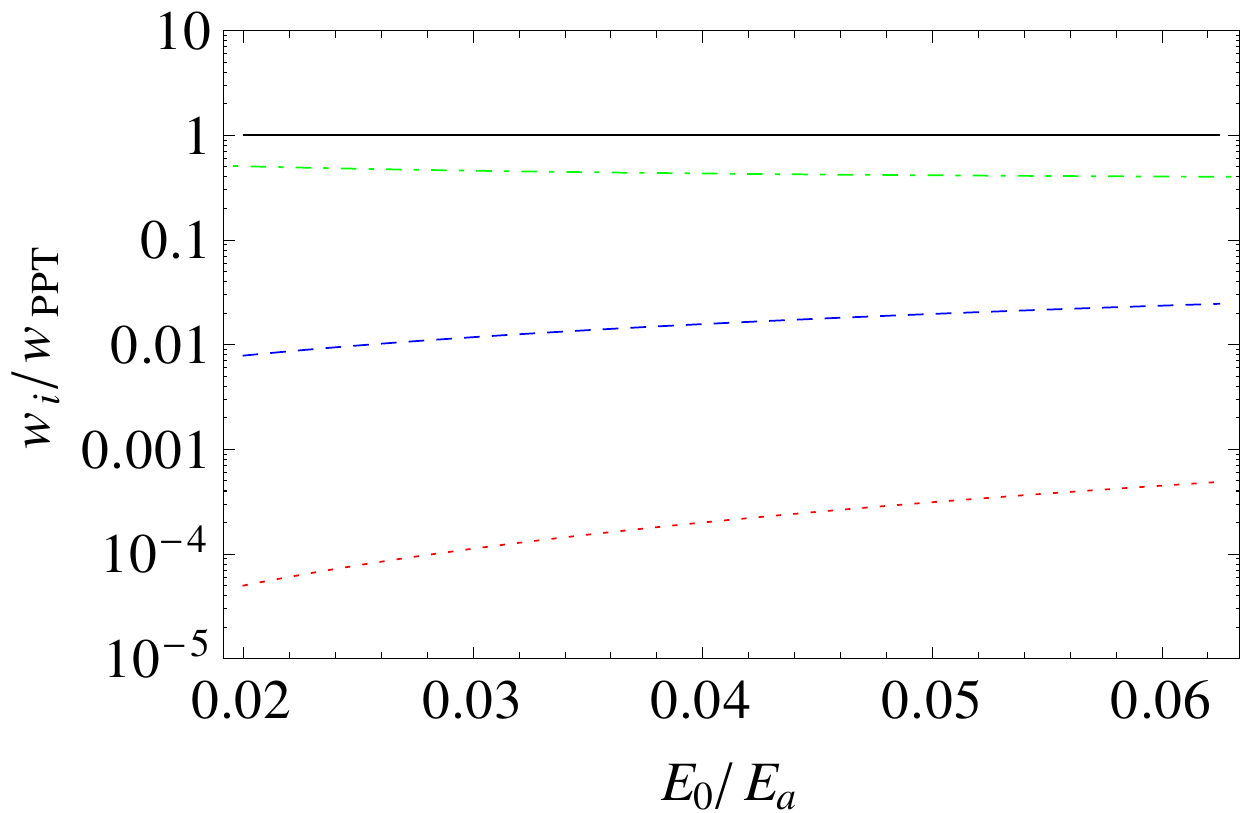}
       \caption{The ratio of the total ionization rate derived in the SFA with respect to the PPT rate vs the parameter $E_0/E_a$: (black, solid) in the Coulomb-corrected SFA in length gauge, (green, dash-dotted) in the Coulomb-corrected SFA in velocity gauge for $\gamma=0.1$, (blue, dashed) standard SFA with a Coulomb-potential, (red, dotted) standard SFA with a zero-range potential.}
    \label{fig3}
  \end{center}
\end{figure}

In the previous sections we have calculated the ionization of a hydrogen-like system in a strong linearly polarized laser field using the Coulomb-corrected SFA in length and velocity gauge.  In Fig.~\ref{fig3} we  compare the total ionization rates in these approximations with the PPT ionization rate for different values of $E_0/E_a$. For comparison also the rates in the standard SFA are presented using a short-range potential and a Coulomb potential.
All  approximations show the same qualitative behavior, but the absolute values of the rates differ significantly. 
The Coulomb-corrected SFA increases the ionization rate by several orders of magnitude. This is in accordance with the intuitive picture that the Coulomb-potential lowers the tunneling barrier and therefore facilitates tunneling. Further, it should be mentioned that the Coulomb-correction is only depending on $E_0/E_a$, but not, e.g., on $I_p$ or $\omega$.

Thus, from the results of this and the previous sections one can conclude that the Coulomb-corrected SFA shows a good agreement with the PPT theory only in length gauge. This is a message that should be taken into account in the generalization of the Coulomb-corrected SFA into the relativistic domain.

\section{Conclusion}\label{Conclusion}

We have applied the Coulomb-corrected SFA for ionization of hydrogen-like systems in a strong linearly polarized laser field. The nonrelativistic regime is considered to show how this approximation works and how to use the developed procedure for a further generalization of the approximation into the relativistic domain. The applied Coulomb-corrected strong-field approximation incorporates the eikonal-Volkov wave function for the description of the electron continuum dynamics. The latter is derived in the WKB approximation taking into account the Coulomb field of the atomic core perturbatively in the phase of the WKB wave function, i.e., in physical terms, the disturbance of the electron energy by the Coulomb field is assumed to be smaller with respect to the electron energy in the laser field.   We have derived an analytical expression for the ionization amplitude within the Coulomb-corrected SFA in length and velocity gauge. A simple expression for the amplitude is obtained when using the length gauge which is due to the fact that the Coulomb correction factor (ratio of the Coulomb corrected amplitude to the standard SFA one) in this gauge cancels the factor of the electric-dipole interaction Hamiltonian in the matrix element. Moreover, a Coulomb correction factor  coinciding with that derived within the PPT theory is obtained. The differential and total ionization rates are calculated analytically. The calculated total ionization rate in length gauge is identical to the PPT-rate, while in the velocity gauge it can deviate from the PPT result up to a factor of 2. Taking into account that the PPT-rate provides a good approximation for experimental results, we can conclude that the Coulomb-corrected SFA works successfully in the length gauge. The SFA in different gauges, in fact, corresponds to different partitions of the total Hamiltonian used to develop the SFA \cite{Faisal_2007b}. Therefore, one can conclude that the relativistic  generalization of the Coulomb-corrected SFA, which will be carried out in the next paper of this sequel, should be based on the partition of the total Hamiltonian that in the nonrelativistic limit corresponds to the partition of the length gauge SFA.

\section*{Acknowledgments}

Valuable discussions with C. H. Keitel and C. M\"uller are acknowledged.

\bibliography{strong_fields_bibliography}

\end{document}